\documentclass[11pt,preprintnumbers,aps,amssymb,nofootinbib,amsmath,superscriptaddress,notitlepage]{revtex4-1}
\usepackage{epsfig,epsf,comment}
\usepackage{bm} 
\usepackage{color} 
\usepackage{slashed}
\usepackage{relsize}	
\usepackage{soul} 
\usepackage{hyperref}
\newcommand{\beq}{\begin{equation}}
\newcommand{\beql}[1]{\begin{equation}\label{#1}}
\newcommand{\eeq}{\end{equation}}
\def\bal#1\gal{\begin{align}#1\end{align}}
\newcommand{\ball}[1]{\bal\label{#1}}
\newcommand{\eq}[1]{(\ref{#1})}
\newcommand{\fig}[1]{Fig.~\ref{#1}}
\renewcommand{\sec}[1]{Sec.~\ref{#1}}
\renewcommand{\b}[1]{{\bm #1}} 
\newcommand{\unit}[1]{\hat {{\bm #1}}} 

\newcommand{\e}{\varepsilon}
\newcommand{\aver}[1]{\left\langle #1 \right\rangle}

\begin{document}

\title{Optical manifestations of domains with constant topological charge density}

\author{Evan Stewart}
\author{Kirill Tuchin}

\affiliation{Department of Physics and Astronomy, Iowa State University, Ames, Iowa, 50011, USA}

\date{\today}


\begin{abstract}

Domains of finite topological charge density can exist in chiral materials and chiral matter. Spatial and temporal variation of 
 the average topological charge density, represented by the $\theta$-field, induces anomalous currents that are responsible for the chiral magnetic effect, the anomalous Hall effect and other phenomena that are intimately related to the chiral anomaly. We consider domains with constant average topological charge density. We argue that even though the Maxwell equations in the bulk are not altered, the chiral anomaly manifests itself by the way of the  boundary conditions. This is illustrated by several examples. The first example deals with the refraction of plane electromagnetic wave on a surface of a constant-$\theta$ domain. We derive the modified Fresnel equations and discuss the effect of the chiral anomaly on the amplitude and polarization of the reflected and transmitted waves. In particular, we argue that the Brewster's angle is sensitive to the value of $\theta$. In the second example we compute the spectrum of the transition radiation at high frequencies and show that it is enhanced at finite $\theta$.

\end{abstract}

\maketitle

\section{Introduction}\label{sec:i}

The chiral anomaly \cite{Adler:1969gk,Bell:1969ts} plays an important role in materials containing chiral fermions. At finite topological charge density they exhibit a number of novel $P$ and $CP$-odd electromagnetic effects \cite{Kharzeev:2015znc}. The topologically non-trivial background has different sources: in hot nuclear matter it is created by the random sphaleron-mediated  transitions between different QCD vacua \cite{Zhitnitsky:2012ej,Kharzeev:2007tn}, in Weyl and Dirac semimetals it emerges in external parallel electric and magnetic fields \cite{Li:2014bha},  in the cosmological models it is due to the presence of the axion field \cite{Marsh:2015xka}. 

The topological charge density $q(x)$ is a microscopic quantity that has a complicated space-time structure. However, in many applications one is only interested in macroscopic properties of materials. The macroscopic theory emerges upon averaging the microscopic quantities over the small domains that contain a macroscopic number of particles. 
The macroscopic quantity that we employ to describe the topological charge density is the $\theta$-field defined as $\theta(x)= -\aver{q(x)}M^{-4}$, where $M$ is a phenomenological parameter of mass\textcolor{blue}{-}dimension one.  Interaction of the electromagnetic field with the domains of finite $\theta$ can be described by adding to the QED Lagrangian the  axion-photon coupling term \cite{Wilczek:1987mv,Carroll:1989vb, Sikivie:1984yz}
\ball{i1}
\mathcal{L}_A =-\frac{c_A}{4}\theta F_{\mu\nu}\tilde F^{\mu\nu}\,,
\gal
where $c_A$ is the QED anomaly coefficient \cite{Adler:1969gk,Bell:1969ts}.  Since $\theta$ depends on $x$, Eq.~\eq{i1} cannot be rewritten as a total derivative and removed from the Lagrangian. Instead, it contributes the spatial and the temporal derivatives of $\theta$ to the modified Maxwell equations:
\begin{subequations}\label{i3}
\bal
&\partial_\mu (F^{\mu\nu}+c_A \tilde F^{\mu\nu}\theta) = j^\nu\,,\label{i4}\\
&\partial_\mu \tilde F^{\mu\nu}= 0\,,\label{i5}
\gal
\end{subequations}
where $\tilde F^{\mu\nu}$ is a dual field. 

We assume that the topological charge density is induced by the external sources, so that $\theta$ is an external non-dynamical field.\footnote{In a fully dynamical theory one includes the kinetic and self-action terms of the $\theta$-field. This yields a macroscopic version of the axion electrodynamics.} Moreover, we assume that $\theta$ is a slowly-varying function of space and time so that at the leading order in derivatives we can write  $\theta(x)\approx \theta(0)+ x\cdot \partial\theta(0)$. Its time derivative  can then be interpreted as the axial chemical potential \cite{Fukushima:2008xe,Kharzeev:2009fn,Kharzeev:2009pj}, while its spacial derivatives (times $c_A$) as the splitting of the Weyl nodes in momentum space \cite{Burkov:2015}.  In the infinite medium, i.e.\ far away from the boundaries,  $\theta(0)$ plays no role as the equations \eq{i3} depend only on the derivatives $\partial \theta$. Most previous studies concentrated on the effect of the chiral anomaly on the electromagnetic field through these derivatives and found a plethora of novel effects, the most remarkable of which are  the chiral magnetic effect (CME) 
\cite{Kharzeev:2004ey,Kharzeev:2007jp,Kharzeev:2015znc,Kharzeev:2013ffa,Li:2014bha} and the chiral instability of electromagnetic field \cite{Joyce:1997uy,Boyarsky:2011uy,Akamatsu:2013pjd}.

However, in finite media, such as the $CP$-odd domains in the quark-gluon plasma, boundaries do matter. For example, recently it was shown in \cite{Tuchin:2018rrw} that the CME is generated even at constant $\theta$ in time-dependent external magnetic field. Moreover, if $\theta$ is indeed a slowly-varying function in the bulk, then its discontinuity at the domain surface is the main source of the chiral effects. It is thus worthwhile to  study  electrodynamics in the presence of domains with constant $\theta$. This is the simplest realization of the chiral anomaly in the electromagnetic theory: even though $\theta$ drops out the anomalous Maxwell equations \eq{i3} in the bulk, it emerges in the boundary conditions on the domain surfaces, see Eqs.~\eq{a26} below.

The main goal of this paper is to use the macroscopic electrodynamics with anomalous boundary conditions to study electromagnetic wave refraction and the transition radiation by constant-$\theta$ domains. We argue that these processes can be used as effective experimental tools to investigate the chiral matter/materials.

The paper is organized as follows. We begin \sec{sec:b} by writing down the boundary conditions \eq{a26} for the electromagnetic fields at the boundary of a finite-$\theta$ domain. We then employ them to study the reflection and transmission of plane electromagnetic waves at the flat boundary. The result is the set of generalized Fresnel equations \eq{a17} for the transition and reflection amplitudes of various polarizations. Our main observation is that the $\theta$-domains are optically active even if the index of refraction is the same on the two sides of the boundary. In \sec{sec:c} we compute the Brewster's angle, i.e.\ the angle at which a plane wave linearly polarized in the plane of incidence is not reflected. We observe that  it increases 
with the wave frequency in a way that strongly depends on $\theta$ as indicated in \fig{brewster}. In \sec{sec:f} we consider emission of the transition radiation by an ultra-relativistic electrically charged fermion crossing the boundary between the vacuum and the $\theta$-domain. This problem has been recently solved for inhomogeneous domains ($\partial\theta\neq 0$) in  \cite{Tuchin:2018sqe,Huang:2018hgk}. Here we compute the spectrum \eq{f9} of the radiated photons in the case of constant $\theta$ and argue that it is enhanced as compared with the conventional transition radiation. We draw conclusions in \sec{sec:s}.


\section{The Fresnel equations for a chiral domain}\label{sec:b}

We focus on linear, homogeneous, isotropic, electrically neutral and nonmagnetic chiral matter with complex refractive index $n(\omega)$. The anomalous Maxwell equations \eq{i3}, also referred to as the Maxwell-Chern-Simons equations,  that describe the macroscopic electrodynamics of this matter can be written in the three-dimensional form as 
\begin{subequations}\label{a-1}
\bal
&\b \nabla\cdot \b B=0\,, \label{a3}\\
& \b \nabla\cdot (\b D+\tilde\theta \b B)= 0\,,  \label{a4}\\
& \b \nabla \times \b E= -\partial_t \b B\,,\label{a5}\\
& \b \nabla \times (\b B-\tilde\theta \b E)= \partial_t (\b E+\tilde\theta \b B) \,,\label{a6i}
\gal
\end{subequations}
where $\b D$ is electric displacement and the pseudo-scalar field $\tilde \theta=c_A \theta$ is proportional to the average topological charge density sourced by the non-Abelian fields or some other topological field configurations mentioned in the previous section.   The boundary conditions on the domain wall can be obtained directly  from equations \eq{a-1}. Denoting by $\Delta$ the discontinuity of a field component across the domain wall  one obtains \cite{Sikivie:1984yz}
\begin{subequations}\label{a26}
\bal
&\Delta B_\bot=0\,, \label{a22}\\
& \Delta  (E_\bot+\tilde\theta  B_\bot)= 0\,,  \label{a23}\\
&  \Delta \b E_\parallel= 0\,,\label{a24}\\
& \Delta (\b B_\parallel-\tilde\theta  \b E_\parallel)= 0 \,.\label{a25}
\gal
\end{subequations}
where $E_\bot$, $B_\bot$  and $\b E_\parallel$, $\b B_\parallel$ are components of the electromagnetic field normal and tangential to the domain wall, respectively.  We would like to use Eqs.~\eq{a26} to compute how the refraction of electromagnetic waves is modified at finite constant $\theta$.

\begin{figure}[t]
      \includegraphics[height=6cm]{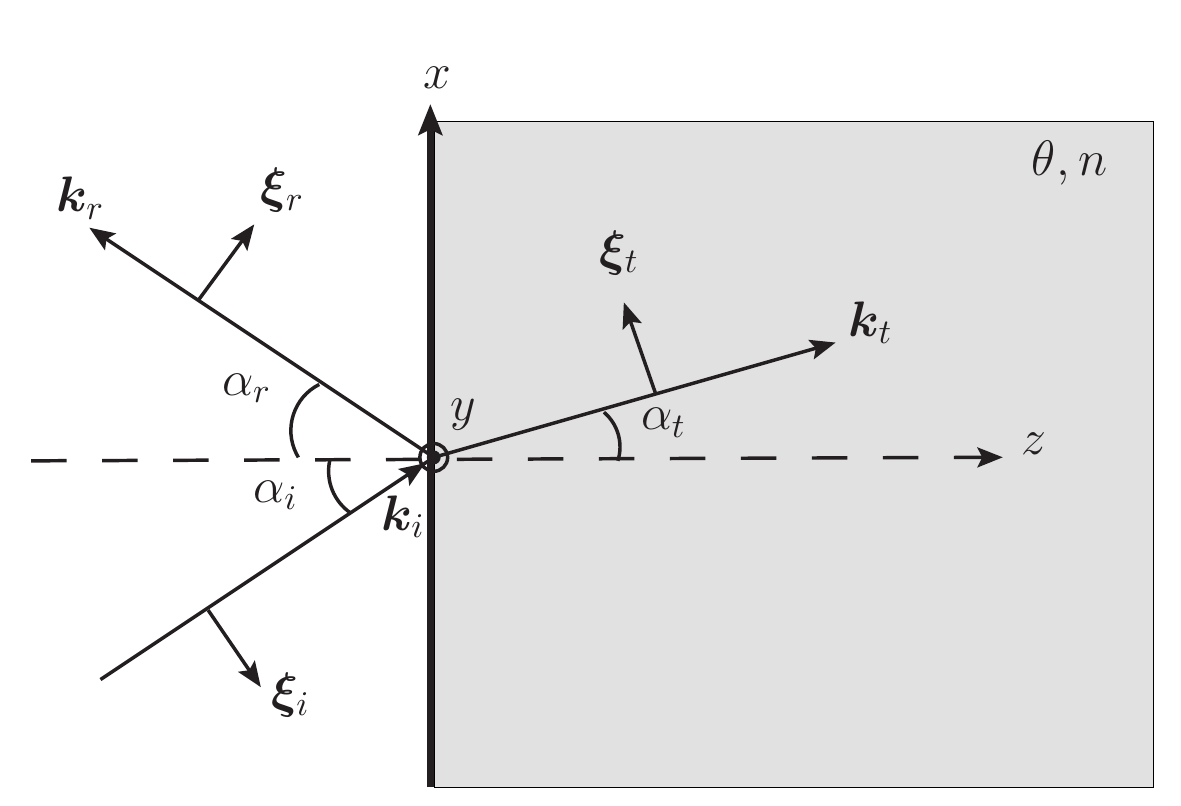} 
  \caption{Scattering of a plane circularly polarized electromagnetic wave off a semi-infinite domain (shaded area at $z\ge 0$) with index of refraction $n$ and constant $\theta$-field at an incident angle $\alpha_i$. The domain boundary is the $xy$-plane. The wave vectors lie in the $xz$-plane and the $y$-axis points at the reader. In the absorptive medium, the refraction index $n$ and the angle $\alpha_t$ are  complex.}
\label{geom}
\end{figure}

Consider the right-hand polarized plane monochromatic electromagnetic wave incident from vacuum ($z<0$) on the flat boundary of a semi-infinite domain ($z>0$) with constant $\theta$-field and constant $n$, see \fig{geom}.\footnote{A similar problem was considered by Hayata \cite{Hayata:2017tbr}. However, unlike the present study, he focused on the effects due to inhomogeneity of $\theta$.} It is convenient to choose the radiation gauge in which the scalar potential vanishes and the vector potential is divergenceless. The vector potential outside the domain is then given by a superposition of the incident and reflected waves, while the vector potential inside the domain is given by the transmitted wave:
\begin{subequations}\label{a0}
\bal
\b A_1&= A_i^+\b \epsilon_i^+ e^{i\b k_i\cdot \b r}+ A_r^+\b \epsilon_r^+e^{i\b k_r\cdot \b r}+A_r^-\b \epsilon_r^-e^{i\b k_r\cdot \b r}\,, \quad z<0\,,\label{a1}\\
\b A_2&= A_t^+\b \epsilon_t^+e^{i\b k_t\cdot \b r}+A_t^-\b \epsilon_t^-e^{i\b k_t\cdot \b r}\,, \quad z>0\,,\label{a2}
\gal
\end{subequations}
where the superscripts $\pm$ refer to the right and left-hand polarizations and we omitted the common time dependence factor $e^{-i\omega t}$. Since $\theta$ is constant at every point except at the boundary, it drops out the equations of motion \eq{a-1}. In particular, the dispersion relations are 
$k_i=k_r=\omega$,  $k_t= n\omega$. Similarly, the continuity of the phase factors across the boundary requires that $k_{ix}=k_{rx}=k_{tx}$, giving the Snell's law $\sin\alpha_i= \sin\alpha_r= n\sin\alpha_t$.
The effect of the finite $\theta$ appears only in the modified Fresnel equations. 

In the reference frame depicted in \fig{geom} the polarization vectors are given by 
\ball{a6}
\b\epsilon_a^\pm=\frac{\unit \xi_a\pm i\unit y}{\sqrt{2}}\,,\quad a=i,r,t\,.
\gal
The circularly polarized plane waves are eigenvectors of the curl operator by virtue of the identity 
\ball{a7}
i\unit k_a\times \b \epsilon_a^\pm = \pm \b \epsilon_a^\pm\,.
\gal
Therefore, the magnetic field corresponding to \eq{a1},\eq{a2} is given by 
\begin{subequations}\label{a8}
\bal
\b B_1&= k_iA_i^+\b \epsilon_i^+ e^{i\b k_i\cdot \b r}+ k_rA_r^+\b \epsilon_r^+e^{i\b k_r\cdot \b r}-k_rA_r^-\b \epsilon_r^-e^{i\b k_r\cdot \b r}\,, \quad z<0\,,\label{a9}\\
\b B_2&= k_tA_t^+\b \epsilon_t^+e^{i\b k_t\cdot \b r}-k_tA_t^-\b \epsilon_t^-e^{i\b k_t\cdot \b r}\,, \quad z>0\,.\label{a10}
\gal
\end{subequations}
The electric field $\b E= i\omega \b A$ is simply proportional to the vector potential. 

Projecting  the boundary conditions \eq{a24} and \eq{a25} on $x$ and $y$ axes yields
\begin{subequations}\label{a11}
\bal
\cos\alpha_i(A_i^+-A_r^+-A_r^-)&= \cos\alpha_t(A_t^++A_t^-)\,,\label{a12}\\
A_i^++A_r^+-A_r^-&= A_t^+-A_t^-\,,\label{a13}\\
\cos\alpha_i(A_i^+-A_r^++A_r^-)&= n\cos\alpha_t(A_t^+-A_t^-)-i\tilde \theta \cos\alpha_t(A_t^++A_t^-)\,,\label{a14}\\
A_i^++A_r^++A_r^-&=n( A_t^++A_t^-)-i\tilde \theta(A_i^+-A_t^-)\,,\label{a15}
\gal
\end{subequations}
The other two boundary conditions \eq{a23} and \eq{a25} do not contain any new information.  Solution to \eq{a12}--\eq{a15} reads
\begin{subequations}\label{a17}
\bal
A_t^+&= \frac{\cos\alpha_i(\cos\alpha_t+\cos\alpha_i)(n+1+i\tilde \theta)}
{n(\cos^2\alpha_i+\cos^2\alpha_t)+\cos\alpha_i\cos\alpha_t(n^2+1+\tilde \theta^2)}A_i^+\,,\label{a18}\\
A_r^-&= \frac{\cos\alpha_t\cos\alpha_i[n^2-(1+i\tilde\theta)^2]}
{n(\cos^2\alpha_i+\cos^2\alpha_t)+\cos\alpha_i\cos\alpha_t(n^2+1+\tilde \theta^2)}A_i^+\label{a19}\\
A_t^-&=\frac{(\cos\alpha_t-\cos\alpha_i)(n-1-i\tilde\theta)\cos\alpha_i}
{n(\cos^2\alpha_i+\cos^2\alpha_t)+\cos\alpha_i\cos\alpha_t(n^2+1+\tilde \theta^2)}A_i^+\,,\label{a20}\\
A_r^+&= \frac{(\cos^2\alpha_i-\cos^2\alpha_t)n}{n(\cos^2\alpha_i+\cos^2\alpha_t)+\cos\alpha_i\cos\alpha_t(n^2+1+\tilde \theta^2)}A_i^+\,,\label{a21}
\gal
\end{subequations}
where $\cos\alpha_t= \sqrt{1-n^{-2}\sin^2\alpha_i}$.
In general, the polarization of the transmitted and reflected waves is elliptical, which is of course also true if $\theta=0$. However, at finite $\theta$ this holds even for the linearly polarized incident wave except in special cases discussed below. 

If the incident wave is left-hand polarized, the corresponding amplitudes can be obtained from \eq{a17} by noting that $\b E$ and $\b B$ are polar and axial vectors respectively. Hence applying the parity transformation to the boundary conditions \eq{a26} is equivalent to replacing $\theta\to -\theta$ and $A_a^\pm\to A_a^\mp$ ($a=i,r,t$) in the Fresnel equations \eq{a17}. 

Consider now several particular cases. In the limit $\theta=0$ one recovers the conventional Fresnel equations. In particular, for the normal incidence $\alpha_i=0$ the amplitudes are
\ball{b1}
A_t^+= \frac{2}{n+1}A_i^+\,,\quad A_r^-= \frac{n-1}{n+1} A_i^+\,,\quad A_t^-&= A_r^+=0\quad \text{if}\,\, \theta=\alpha_i=0\,.
\gal

If the media on the two sides of the boundary have the same index of refraction, but different values of $\theta$, e.g.\ the two $\theta$-vacua in QCD, the boundary is still reflective.  In particular, setting  $n=1$ implies that   $\alpha_i= \alpha_r=\alpha_t$ and the Fresnel equations  \eq{a17} reduce to (at any $\alpha_i$)
\bal
A_t^+&= \frac{2}{2-i\tilde\theta}A_i^+\,,\quad  A_r^-=-\frac{i\tilde\theta}{2-i\tilde\theta}A_i^+\,, \quad A_t^-= A_r^+=0\quad \text{if}\,\, n=1\,.\label{b2}
\gal
We observe that (i) the amplitudes of the transmitted and reflected waves do not depend on the angle of incidence, 
 (ii) a material with finite $\theta$ reflects electromagnetic waves with the effective index of refraction $n_\text{eff}=1-i\tilde\theta$ (as can be realized by comparing \eq{b1} and \eq{b2}), and (iii) the polarization of the transmitted wave is the same as the  polarization of the incident wave. In other words, at $n=1$, $\theta\neq 0$  the circular polarization is preserved, in contrast to the $n>1$, $\theta=0$ case where the linear polarization is preserved. It follows from \eq{b2} that an electromagnetic wave traversing domains with different values of $\theta$, but the same refractive index, is reflected even at normal incidence. Experimentally, Eqs.~\eq{b2} describe scattering of high frequency waves in which case the index of refraction is close to unity. We will return to this point in more detail in the following sections. In Appendix~A we record the result for the refraction of a right-hand polarized wave normally incident at a thin film.

\section{Brewster's angle}\label{sec:c}

\begin{figure}[t]
      \includegraphics[height=5cm]{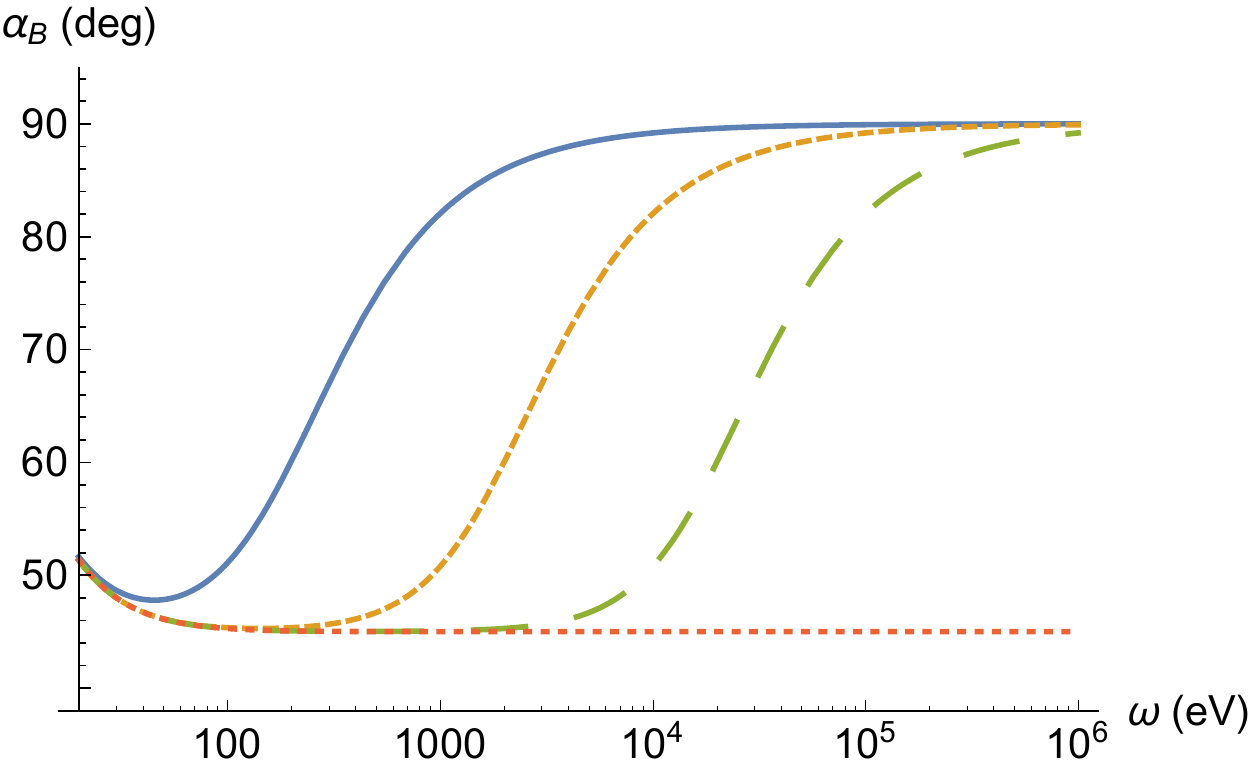} 
  \caption{Brewster's angle as a function of the wave frequency for a typical value of the plasma frequency $\omega_\text{pl}=10$~eV at different values of  $\tilde\theta$ = $10^{-1}$, $10^{-2}$, $10^{-3}$ (left to right).  The dotted line corresponds to $\theta=0$. }
\label{brewster}
\end{figure}

At a certain angle of incidence, known as the Brewster angle, there is no reflection of the incident wave component linearly polarized in $xz$-plane. To determine the dependence of this angle on $\theta$, consider the linearly polarized  incident wave. It can be written as a superposition of the right and left-hand polarized waves as (throughout this section we omit the phase factors for brievity)
\ball{c1}
\b A_i = A_i^+\b\epsilon_i^++ \mathcal{A}_i^-\b\epsilon_i^-= \sqrt{2}A_i^+
\left\{ 
\begin{array}{cc}
\unit \xi_i  & \text{if}\,\, A_i^+= \mathcal{A}_i^- \\
\unit y & \text{if}\,\, A_i^+=- \mathcal{A}_i^-
\end{array}
\right.
\,,
\gal
The calligraphic font refers to the amplitude of the land-hand polarized component to distinguish it from the results of the previous section (recall that Eqs.~\eq{a17} are written for the right-hand polarized incident wave). The vector potential of the reflected wave is
\ball{c3}
\b A_r= (A_r^+  + \mathcal{A}_r^+) \b\epsilon_r^+ +(A_r^-+ \mathcal{A}_r^-)\b\epsilon_r^-\,.
\gal
Substituting \eq{a19},\eq{a21} into \eq{c3} and recalling that, as explained in the previous section, $ \mathcal{A}_r^\pm(\theta) = A_r^\mp(-\theta)$ we obtain for the incident wave polarized in $xz$ plane (i.e.\ $A_i^+= \mathcal{A}_i^-$):
\ball{c5}
\b A_r= \sqrt{2}A_i^+ & \left\{ \unit \xi_r 
\frac{n(\cos^2\alpha_i-\cos^2\alpha_t)+\cos\alpha_i\cos\alpha_t(n^2-1+\tilde\theta^2)}{n(\cos^2\alpha_i+\cos^2\alpha_t)+\cos\alpha_i\cos\alpha_t(n^2-1+\tilde\theta^2)} \right. \nonumber\\
&\unit y \left.\frac{2i\tilde\theta\cos\alpha_i\cos\alpha_t}{n(\cos^2\alpha_i+\cos^2\alpha_t)+\cos\alpha_i\cos\alpha_t(n^2-1+\tilde\theta^2)} 
\right\}
\gal
The in-plane component of the reflected wave vanishes when $\alpha_i=\alpha_B$ where 
\ball{c6}
\tan^2\alpha_B= \frac{(n^2-1+\tilde\theta^2)\left(n^2-1+\tilde\theta^2+\sqrt{(n^2+1)^2+\tilde\theta^2[\tilde\theta^2+2(n^2-1)]} \right)}{2(n^2-1)}\,.
\gal
In \fig{brewster} we plotted $\alpha_B(\omega)$ for several values of $\tilde\theta$. One can see that the Brewster's angle is quite sensitive to $\tilde\theta$. 

There are two limiting cases: if $|\tilde\theta|\ll 1$ at fixed  $n$ then
\ball{c7}
\tan\alpha_B\approx n + \frac{n^3}{n^4-1}\tilde \theta^2\,.
\gal
The first term on the right-hand side is the conventional result for $\theta=0$. The second term is a correction that implies that the Brewster's angle increases with $\tilde\theta$. At high frequencies another limit can be achieved: keeping $\tilde\theta$ fixed at taking  $n-1\ll 1$ we have
\ball{c8}
\tan\alpha_B\approx \frac{\tilde\theta}{2\sqrt{n-1}}\sqrt{\tilde\theta^2+\sqrt{4+\tilde\theta^4}}
\gal
In this case the Brewster's angle becomes close to $\pi/2$. Since at large frequencies $n-1$ is proportional to $1/\omega^2$,  we derive that this limit as approached as $\pi/2-\theta_B\propto 1/\omega$.

\section{Transition radiation}\label{sec:f}

The conventional transition radiation is emitted when a fast charged particle, i.e.\ a particle moving with energy much greater than the medium ionization energy, crosses the boundary between the two media having different dielectric constants \cite{Ginzburg:1945zz,Ginzburg-Tsytovich,Baier:1998ej,Schildknecht:2005sc}.  Its main application is the tomographic investigation of the medium electromagnetic properties. The transition radiation originates  from the difference of the photon wave function on the two sides of the boundary. 
In chiral matter with $\partial\theta\neq 0$ the photon dispersion relation is modified due to the chiral anomaly. As a result, when a fast charged particle crosses the boundary between the chiral matter and vacuum it emits the chiral transition radiation which was studied in  \cite{Tuchin:2018sqe,Huang:2018hgk}. In this section we argue that the chiral transition radiation is emitted even if  $\theta$ is constant everywhere except at the boundary. 

\begin{figure}[t]
      \includegraphics[height=3cm]{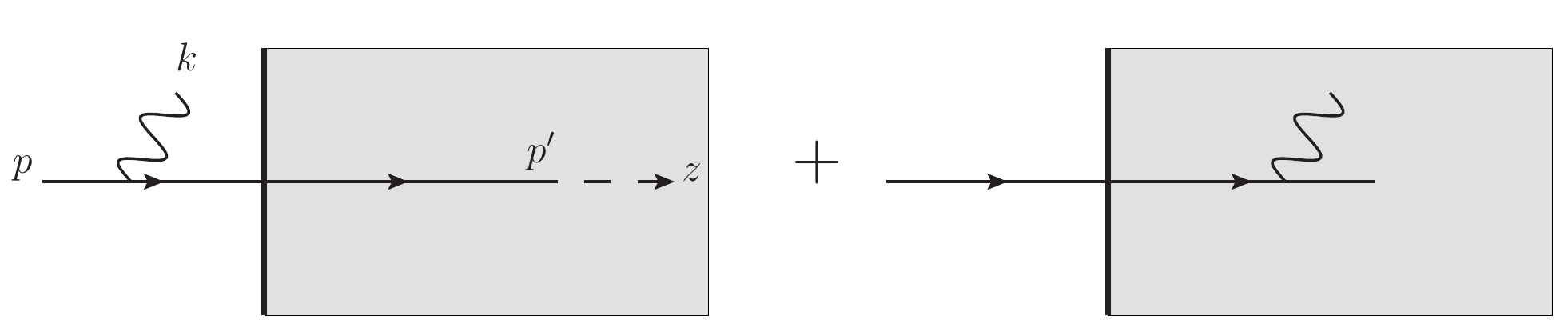} 
  \caption{Diagrams contributing to the photon radiation by a charged particle crossing the boundary. }
\label{geom3}
\end{figure}

Consider a charged ultra-relativistic fermion moving along the $z$-axis as shown in \fig{geom3}. We assume that 
emitted photon energy $\omega$ is much larger than the plasma frequency of the medium $\omega_\text{pl}$,  which allows writing  the photon dispersion relation as $\omega^2= \b k^2+\omega_\text{pl}^2$. The corresponding index of refraction is close to unity: $n-1= \omega_\text{pl}^2/2\omega^2$. If  the photon energy is so large that $n-1\ll |\tilde \theta|$, one can neglect the deviation of the refractive index from unity. In this case the scattering amplitudes are given by Eqs.~\eq{b2}. The fact that they are independent of the photon momentum makes the calculation especially simple.

In the high energy limit the wave function of the right-hand polarized photon, normalized to one particle per unit volume, reads
\ball{f1}
\b A^+ = \frac{1}{\sqrt{2\omega}}\left\{ \left[\b\epsilon_i^+e^{i\b k_i\cdot \b r}-\frac{i\tilde \theta}{2-i\tilde \theta}\b\epsilon_r^-e^{i\b k_r\cdot \b r}\right]\eta(-z)
+\frac{2}{2-i\tilde \theta}\b\epsilon_t^+e^{i\b k_t\cdot \b r}\eta(z)
\right\}e^{-i\omega t} \quad \text{if}\,\, \omega\gg \frac{\omega_\text{pl}}{\sqrt{\tilde \theta}}\,,
\gal
 where $\eta$ is a step function and the normalization volume is set to unity here and in \eq{f3}. The wave function of the left-hand polarized photon $\b A^-$ can be obtained by complex conjugating the polarization vectors and the amplitudes in front of them.  Let $\b k_\bot$ denote the photon momentum perpendicular to the the fermion direction $\unit z$. Note that $k_{ix}=k_{rx}=k_{tx}=k_\bot$ as indicated in \sec{sec:b}.\footnote{The subscript $\bot$ here and in Eqs.~\eq{a26} are not to be confused.} As for the longitudinal components, they are given by $k_{iz}=-k_{rz}\equiv k_z$ and $k_{tz}= \sqrt{n^2k_{z}^2+(n^2-1)k_\bot^2}\approx k_z$. The quantization of the electromagnetic field in the presence of the semi-infinite dielectric material was discussed in \cite{Carniglia:1971te,BialynickiBirula:1972jx}.
 
 The scattering matrix element for the photon emission at the leading order is 
 \ball{f2}
 S= -ie Q\int \bar \psi_{p'} \gamma^\mu \psi_p A^*_\mu d^4x\,,
 \gal
  where $Q$ is the fermion electric charge in units of electron charge $e$ and $\psi_p$ is its wave function, which in the ultra-relativistic approximation reads
 \ball{f3}
\psi_p= \frac{1}{\sqrt{2\e}}u(p)  e^{i\e (z- t)}\exp\left\{ i\b p_\bot\cdot \b x_\bot-iz\frac{\b p_\bot^2+m^2}{2\e}\right\}\,,
\gal
where $\e$, $\b p$ are the initial fermion energy and momentum respectively and $\b x_\bot$ is a position vector in the $xy$ plane. The primed variables refer to the final fermion. In the reference frame that we chose $\b p_\bot=0$ .  Taking integrals over time and $x_\bot$ one obtains the matrix element for the right-handed photon emission
\ball{f4}
S=& -ie Q(2\pi)^3\delta(\omega+\e'-\e)\delta(\b k_\bot+\b p'_\bot)\frac{\bar u (p')\slashed{\epsilon^*} u(p)}{\sqrt{8\e\e' \omega}}\nonumber\\
&\times \int_{-\infty}^\infty dz\, e^{i(p_z-p_z')z}\left[ \eta(-z)\left( e^{-ik_zz}-\frac{i\tilde\theta}{2-i\tilde\theta}e^{ik_zz}\right)+\eta(z)\frac{2}{2-i\tilde \theta}e^{-ik_zz}\right]\,.
\gal
The longitudinal momentum transfer in the forward direction is small
\ball{f5}
p_z-p_z'-k_z\approx -\frac{m^2}{2\e}+\frac{{p'}^2_\bot+m^2}{2\e'}+\frac{k_\bot^2+\omega_\text{pl}^2\eta(z)}{2\omega}
=\frac{ k_\bot^2+x^2m^2+(1-x)\omega^2_\text{pl}\eta(z)}{2\e x(1-x)}\,,
\gal
where $x=\omega/\e$. However, the longitudinal momentum transfer in the backward direction is very large $p_z-p_z'+k_z\approx 2\omega$. Since the integrals in \eq{f4} are inversely proportional to the longitudinal momentum transfer, the reflected wave can be neglected in the high-energy limit. Thus the second line in \eq{f4} yields
\ball{f7}
\int_{-\infty}^\infty dz\,\ldots = -\frac{2i\e x(1-x)}{k_\bot^2+x^2m^2}+\frac{2}{2-i\tilde\theta}\,\frac{2i\e x(1-x)}{k_\bot^2+x^2m^2+(1-x)\omega^2_\text{pl}}
\,.
\gal
Evidently, the two terms on the right-hand-side correspond to the two diagrams in \fig{geom3}. The derivation of the photon spectrum from the scattering matrix follows the standard procedure with the result
\ball{f8}
\frac{dN}{d^2k_\bot dx}= &\frac{\alpha Q^2}{2\pi^2 x}\left\{ \left(\frac{x^2}{2}-x+1\right)k_\bot^2+\frac{x^4m^2}{2}\right\}
\nonumber\\
&\times
\sum_\pm \left| \frac{1}{k_\bot^2+x^2m^2}-\frac{2}{2\mp i\tilde\theta}\,\frac{1}{k_\bot^2+x^2m^2+(1-x)\omega^2_\text{pl}}\right|^2\,.
\gal
It is easy to see that the two polarizations contribute equally to the spectrum that reads
\ball{f9}
\frac{dN}{d^2k_\bot dx}= &\frac{\alpha Q^2}{\pi^2 x}\left\{ \left(\frac{x^2}{2}-x+1\right)k_\bot^2+\frac{x^4m^2}{2}\right\}\frac{1}{k_\bot^2+x^2m^2+(1-x)\omega^2_\text{pl}}\nonumber\\
&
\times \left[  \frac{(1-x)^2\omega_\text{pl}^4}{(k_\bot^2+x^2m^2)^2(k_\bot^2+x^2m^2+(1-x)\omega^2_\text{pl})}
\right.\nonumber\\
&\left.
+\frac{\tilde\theta^2}{4+\tilde\theta^2}\,\left(  \frac{2}{k_\bot^2+x^2m^2+(1-x)\omega^2_\text{pl}}  -\frac{1}{k_\bot^2+x^2m^2}\right) \right]\,.
\gal

\begin{figure}[ht]
      \includegraphics[height=5cm]{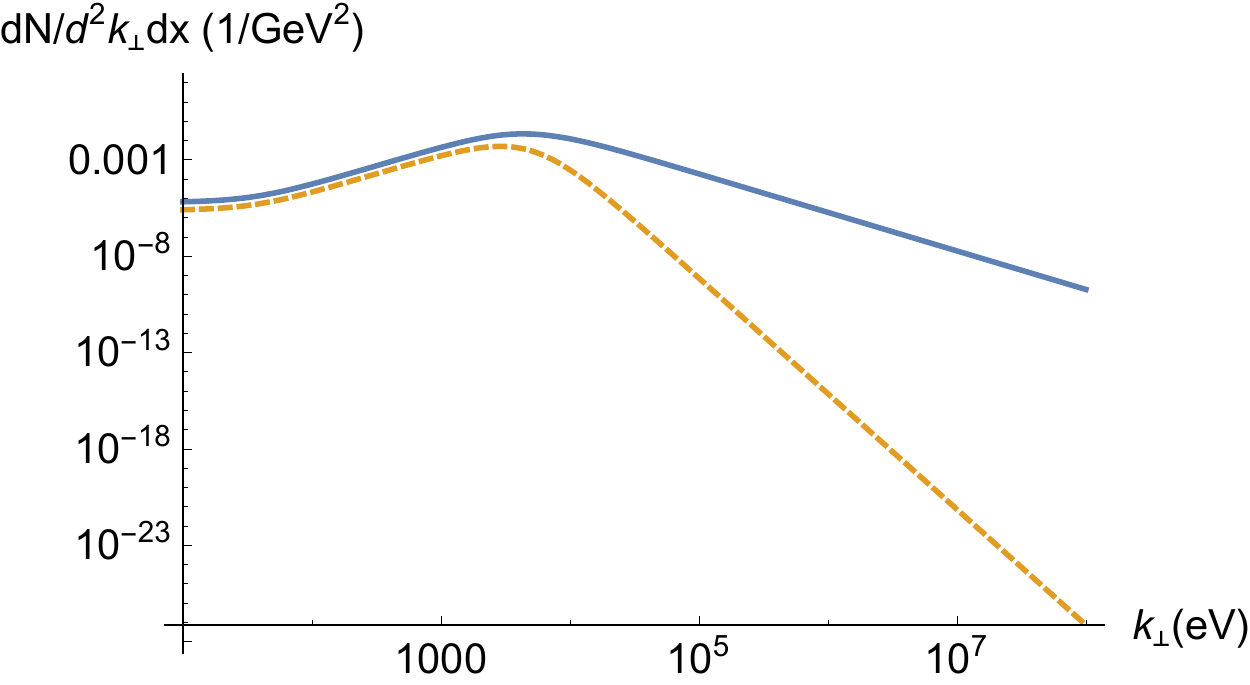} 
  \caption{Transition radiation spectrum for $Q=1$, $m=0.5$~MeV (electron), $\omega_\text{pl}=10$~eV, $x=10^{-2}$, $\tilde\theta= 10^{-5}$. Solid line is the full spectrum, the dashed line is its non-anomalous component. }
\label{spectrum}
\end{figure}

In a particular case  $\theta=0$ one obtains the usual formula for the transition radiation. The $\theta$-independent contribution in \eq{f9} falls off as $\omega_\text{pl}^4/k_\bot^6$ with the photon transverse momentum, whereas the anomalous term only as $\tilde \theta^2/k_\bot^2$. We also note that the contribution of the gradients $\partial\theta$ to the transition radiation, calculated in \cite{Tuchin:2018sqe,Huang:2018hgk} also decreases as $k_\bot^{-6}$. Therefore,  the anomalous term in \eq{f9} dominates at large transverse momenta $k_\bot \gg \omega_\text{pl}/\sqrt{\tilde\theta}$. This is illustrated in \fig{spectrum}.

\section{Summary}\label{sec:s}

Domains of constant topological charge impact the electromagnetic processes through the boundary conditions that depend on the parameter $\tilde\theta= c_A\theta$. We used two examples to demonstrate how this happens and to propose possible avenues for experimental investigation.  

In the first example, we considered the refraction of the electromagnetic wave on flat surface of a domain with finite $\theta$. We derived the modified Fresnel equations \eq{a17} for the amplitudes of  reflected and transmitted waves. At high frequencies, when the refraction index is close to unity, there is still ``topological" refraction due only to the finite value of $\theta$. In this case, the amplitudes of the transmitted and reflected waves do not depend on the angle of incidence and it is the circular polarization that is preserved (not the linear one as in the conventional case $\theta=0$). One can measure the value of $\tilde\theta$ by observing the Brewster's angle as shown in \fig{brewster}.

In the second example we derived the spectrum of  transition radiation for an ultra-relativistic particle crossing the boundary between the vacuum and the finite-$\theta$ domain. We found that at high transverse momenta, photon radiation due to the chiral anomaly dominate over the conventional transition radiation mechanism. Moreover, it also dominates over the chiral transition radiation emitted due to the finite time derivative of $\theta$.  Thus, this represents an opportunity to establish existence of the  $CP$-odd domains of constant topological charge density and measure the magnitude of  $\theta$.

\acknowledgments
  This work  was supported in part by the U.S. Department of Energy under Grant No.\ DE-FG02-87ER40371.

\appendix
\section{Refraction at a thin film}\label{appA}

If a right-hand polarized wave is  normally incident at a thin film of width $d$, then by following the prescription of \sec{sec:b} we obtain for the first boundary
\begin{subequations}
\bal
A_i^++A_r^-=a_t^++a_r^-\,, \\
A_i^+-A_r^-=a_t^+(n-i \tilde\theta n)+a_r^-(n+i\tilde \theta n)\,,
\gal
\end{subequations}
where $a_r^-$ and $a_t^+$ denote the amplitudes of the left and right-propagating waves inside the film.

For the second boundary we obtain
\begin{subequations}
\bal
a_t^+\exp(id\omega n)+a_r^-\exp(-id\omega n)=A_t^+\exp(i\omega d)\,, \\
a_t^+(n-i\tilde \theta n)\exp(id\omega n)+a_r^-(n+i \tilde\theta n)\exp(-id\omega n)=A_t^+\exp(i\omega d)\,.
\gal
\end{subequations}

The amplitudes of the non-vanishing transmitted and reflected waves are
\begin{subequations} \label{b10}
\bal 
A_t^+&= \frac{4n\exp(id\omega (n-1))}
  {1+n(2+n+n\tilde\theta^2)-\exp(2id\omega n)(1+n(-2+n+n\tilde\theta^2))}A_i^+\,,\label{b11}\\
A_r^-&= \frac{(1-n(n-2i\tilde\theta+n\tilde\theta^2))\sin(d\omega n)}
  {2i n\cos(d\omega n)+(1+n^2(1+\tilde\theta^2))\sin(d\omega n)}A_i^+\,.\label{b12}
\gal
\end{subequations}
At $\theta=0$ these equations reduce to the known result \cite{Born:1999ory}. At high frequencies $n=1$ and Eqs.~\eq{b10} reduce to
\begin{subequations} \label{b15}
\bal 
A_t^+&= \frac{4}
  {4+\tilde\theta^2(1-\exp(2id\omega ))}A_i^+\,,\label{b16}\\
A_r^-&= \frac{\tilde\theta(\tilde\theta-2i))\sin(d\omega)}
  {2i \cos(d\omega)+(2+\tilde\theta^2)\sin(d\omega)}A_i^+\,.\label{b17}
\gal
\end{subequations}


\end{document}